\documentclass[11pt,a4paper,twocolumn]{article}

\usepackage[margin=1in]{geometry}
\usepackage{amsmath,amssymb,amsthm}
\usepackage{graphicx}
\usepackage{tikz}
\usetikzlibrary{calc,positioning,shapes.geometric,arrows.meta,decorations.pathreplacing,backgrounds,3d,perspective,fit}
\usepackage{booktabs}
\usepackage{multirow}
\usepackage{algorithm}
\usepackage{algpseudocode}
\usepackage{xcolor}
\usepackage{enumitem}
\usepackage{caption}
\usepackage{subcaption}
\usepackage{float}
\usepackage{cite}
\usepackage{microtype}
\usepackage{hyperref}
\allowdisplaybreaks
\hypersetup{
    colorlinks=true,
    linkcolor=blue!60!black,
    citecolor=blue!40!black,
    urlcolor=blue!60!black
}
\definecolor{lorentzblue}{RGB}{41,98,255}
\definecolor{bm25orange}{RGB}{255,145,0}
\definecolor{hybridgreen}{RGB}{0,170,80}
\title{\textbf{Real-Time Hybrid Retrieval in Hyperbolic Space for Retrieval-Augmented Generation on Edge Devices}}
\author{Aradhya Chakrabarti\\
        \small International Institute of Information Technology\\
        \small Bangalore, India\\
        \small \texttt{aradhya.chakrabarti@iiitb.ac.in}}
\date{}
\newcommand{\keywords}[1]{\vspace{4pt}\noindent{\small\textbf{Keywords:} #1}}

\begin{document}
\maketitle

\begin{abstract}
This paper presents a hybrid document retrieval system designed for retrieval-augmented generation (RAG) that operates entirely within the Lorentz model of hyperbolic geometry. Unlike conventional dense retrievers confined to Euclidean space, this system projects pretrained word embeddings into hyperbolic space through a learned HyTE-H transformation, whose exponential volume growth suits the hierarchical organization of natural language. Documents are segmented into overlapping chunks, indexed by their Lorentz embeddings, and retrieved through a two-stage pipeline that first applies BM25 lexical scoring, then re-ranks candidates using Lorentzian inner-product similarity. A tunable parameter $\alpha$ blends the BM25 score with the hyperbolic similarity score. The system was evaluated on five datasets from the BEIR benchmark suite, SciFact, NFCorpus, ArguAna, SciDocs, and FiQA, achieving NDCG@10 scores of 0.654, 0.304, 0.342, 0.150, and 0.217 respectively with word embeddings alone, without fine-tuned neural encoders or cross-attention rerankers. The system supports real-time indexing of user-supplied documents and resource-efficient querying over tens of thousands of moderately sized documents, so hyperbolic retrieval can run on edge devices at interactive latencies.
\end{abstract}


\section{Introduction}

Dense retrieval is the first stage of modern retrieval-augmented generation~\cite{lewis2020retrieval}, and the evidence it selects determines the quality of every generated response. When a user queries a RAG system, the retriever must pick a few relevant passages from a potentially vast corpus within tens of milliseconds, and mistakes here degrade the quality of the generated answer~\cite{fan2024ragsurvey}. Yet the geometry of most deployed retrievers, Euclidean vector space, was inherited from general-purpose neural network architectures rather than chosen for how language is organized.

Natural language is hierarchical, and Euclidean spaces represent that structure poorly. For example, ``machine learning'' branches into ``deep learning'' and ``natural language processing'', and these subdivide further into ``transformer attention mechanisms'' and ``contrastive pretraining objectives''. In a Euclidean embedding space, a ball of radius $r$ has volume $O(r^d)$. An exponentially growing number of distinct hierarchical branches therefore has to fit into a region whose volume grows only polynomially. This crowding phenomenon~\cite{radovanovic2010hubs} makes semantically distinct but hierarchically related documents look close together in embedding space, which hurts retrieval precision~\cite{reimers2021sentence}.

Hyperbolic geometry, specifically the Lorentz hyperboloid model, avoids this crowding problem. In a space of constant negative curvature $-1/K$, the volume of a ball of radius $r$ grows exponentially rather than polynomially, matching the branching of natural hierarchies. In this geometry, the distance from the origin measures specificity: generic concepts sit near the centre, and specialised concepts sit near the boundary~\cite{nickel2017poincare,mishne2022lorentz}. This radial organization comes from the geometry itself, not from how the embeddings were trained. HypRAG~\cite{madhu2025hyprag} shows that this improves context relevance and answer faithfulness in RAG systems.

The system described in this paper is a complete implementation of hyperbolic document retrieval for edge devices. It has four components:

\begin{enumerate}
    \item \textbf{HyTE-H projection module:} Maps BGE-small word embeddings into the Lorentz hyperboloid through a two-layer nonlinear transformation trained contrastively on MS MARCO passage pairs, preserving hierarchical structure without retraining the underlying language model.
    
    \item \textbf{Lorentz index:} Stores document chunk embeddings as a flat array of float32 Lorentz vectors, with top-$k$ retrieval by exhaustive inner-product maximisation.
    
    \item \textbf{Hybrid BM25-Lorentz reranking pipeline:} Performs a two-stage retrieval in which BM25 lexical scoring identifies an initial candidate set, after which Lorentzian similarity re-ranks the candidates, with the two scores blended by $\alpha$.
    
    \item \textbf{End-to-end runtime:} Packages the entire pipeline: word embedding lookup, hyperbolic projection, chunking, indexing, BM25 scoring, and Lorentz retrieval, into a self-contained C++ shared library.
\end{enumerate}

The rest of this paper covers related work in hyperbolic representation learning (Section~\ref{sec:related}), the design of the system (Section~\ref{sec:methodology}), retrieval results on five BEIR datasets (Section~\ref{sec:results}), and the engineering challenges of deploying hyperbolic retrieval on edge devices (Section~\ref{sec:conclusion}).

\section{Related Work}\label{sec:related}

\subsection{Hyperbolic Representation Learning}

The use of hyperbolic geometry for machine learning traces back to Nickel and Kiela's seminal work on Poincar\'{e} embeddings~\cite{nickel2017poincare}, which demonstrated that symbolic data with latent hierarchical structure, WordNet taxonomies, phylogenetic trees, and lexical entailment graphs, could be embedded into the Poincar\'{e} ball with substantially lower distortion than Euclidean alternatives. They showed that the Poincar\'{e} ball's exponential volume growth allows tree-like structures, whose number of nodes grows exponentially with depth, to be accommodated without the crowding that arises in Euclidean embeddings of the same dimensionality.

Subsequent work by Nickel and Kiela~\cite{nickel2018lorentz} moved from the Poincar\'{e} ball to the Lorentz (hyperboloid) model, which is more numerically stable during Riemannian optimisation. Unlike the Poincar\'{e} ball, whose distance function involves a logarithm of a ratio that approaches zero near the boundary, the Lorentz model expresses geodesic distance through a simple inner product, yielding gradients that remain well-behaved throughout the embedding manifold. Because the Lorentz model is also isometric to the Poincar\'{e} ball (the same geometry in different coordinate charts), it has become the preferred representation for neural hyperbolic learning.

Ganea, B\'{e}cigneul, and Hofmann~\cite{ganea2018hyperbolic} laid the algorithmic foundations for hyperbolic neural networks by deriving hyperbolic analogues of multinomial logistic regression, feed-forward layers, and recurrent architectures within the framework of M\"{o}bius gyrovector spaces on the Poincar\'{e} ball. Their hyperbolic GRU outperformed Euclidean counterparts on textual entailment, showing that neural computation works natively in curved spaces without losing expressivity. Chami et al.~\cite{chami2019hyperbolic} extended this line of work to graph-structured data with HGCN, the first inductive hyperbolic graph convolutional network operating in the Lorentz model with trainable curvature per layer, achieving up to 63\% error reduction on link prediction benchmarks. Subsequent work~\cite{chami2020low} scaled hyperbolic tree embeddings to commodity GPU hardware using floating-point expansion arithmetic and better angular separation.

More recently, HELM~\cite{helm2025} scaled hyperbolic geometry to billion-parameter language models with a mixture-of-curvature experts architecture, where each expert operates in its own curvature space and a router assigns tokens to the best-fitting subspace. The hyperbolic multi-head latent attention and rotary positional encodings introduced in HELM extend these benefits from embeddings to the full transformer stack.

\subsection{Hyperbolic Retrieval and HypRAG}

The most directly relevant prior work is HypRAG~\cite{madhu2025hyprag}, which introduced hyperbolic dense retrieval for RAG with two model variants: HyTE-FH, a fully hyperbolic transformer operating entirely within the Lorentz model, and HyTE-H, a hybrid architecture that projects pretrained Euclidean embeddings into hyperbolic space through a learned transformation. HypRAG's key innovation, and the component this work adopts, is the \textbf{Outward Einstein Midpoint} (OEM), a pooling operator that aggregates token-level points on the hyperboloid.

With a plain Euclidean midpoint, averaging more points pulls the result toward the origin of the hyperboloid and erases the hierarchy encoded in their radial coordinates. The OEM operator counters this through a radial re-weighting mechanism. Given a set of $n$ token-level points $\mathbf{p}_i = (t_i, \mathbf{x}_i)$ on the hyperboloid $\mathbb{H}^d_K$ where $t_i = \sqrt{\|\mathbf{x}_i\|^2 + K}$ is the time coordinate, the OEM-weighted aggregate is computed in two steps. First, the weighted Euclidean sum in the ambient Minkowski space:
\begin{equation}
\tilde{\mathbf{p}} = \frac{\sum_{i=1}^{n} t_i^{P+1} \mathbf{p}_i}{\sum_{i=1}^{n} t_i^{P+1}}
\end{equation}

which gives more weight to points with larger time coordinates (more specific concepts) through the radial weight $t_i^{P+1}$. Since this weighted average generally lies off the hyperboloid, a reprojection step maps it back onto the manifold through radial rescaling:

\begin{equation}
\mathbf{p}_{\text{OEM}} = \sqrt{\frac{K}{-\langle\tilde{\mathbf{p}}, \tilde{\mathbf{p}}\rangle_{\mathcal{L}}}} \;\; \tilde{\mathbf{p}}
\end{equation}

The exponent $P \geq 0$ sets how strongly the deeper (larger $t$) points dominate the aggregate. With $P = 0$, the OEM reduces to projection of the standard Einstein midpoint; with $P = 1$ (our default), the operator creates a mild specificity bias; and as $P \to \infty$, only the single point with the largest time coordinate influences the result, reducing to a max-pooling operation.

HypRAG demonstrated on MTEB and RAG Bench that HyTE-H with OEM achieves up to 29\% gains over Euclidean baselines.

\subsection{Sparse-Dense Hybrid Retrieval}

Purely embedding-based retrievers perform poorly on queries containing rare named entities, numerical values, or highly specific technical terms that may not appear in the pretraining distribution of the encoder~\cite{izacard2022unsupervised}. Sparse lexical retrieval methods, particularly BM25~\cite{robertson2009probabilistic}, complement dense methods by providing exact term matching with strong performance on keyword-heavy queries.

The SPLADE family~\cite{formal2021splade} and ColBERT~\cite{khattab2020colbert} demonstrated that architectures combining sparse lexical signals with dense embeddings could outperform either approach in isolation. Our system follows the same idea within the resource limits of edge devices: BM25 retrieves an initial candidate set, which Lorentz inner-product similarity then re-ranks (Section~\ref{sec:methodology}), with the BM25 stage dominating the cost.

\section{Methodology}\label{sec:methodology}

\subsection{System Architecture}

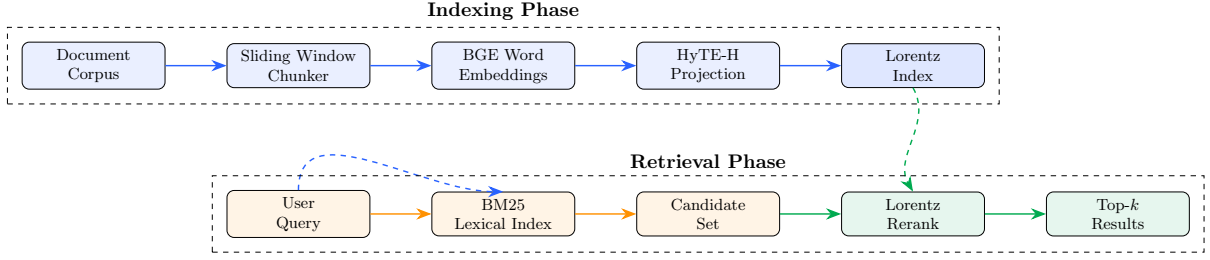
\begin{figure*}[!tb]
\centering
\resizebox{\textwidth}{!}{%
\begin{tikzpicture}[
    node distance=1.2cm,
    box/.style={rectangle, draw, rounded corners, minimum width=2.8cm, minimum height=0.8cm, align=center, font=\footnotesize},
    arrow/.style={-{Stealth[length=3mm]}, thick},
    phase/.style={rectangle, draw, dashed, inner sep=8pt, font=\footnotesize\itshape}
]

\node[box, fill=lorentzblue!10] (docs) {Document\\Corpus};
\node[box, fill=lorentzblue!10, right=of docs] (chunk) {Sliding Window\\Chunker};
\node[box, fill=lorentzblue!10, right=of chunk] (wemb) {BGE Word\\Embeddings};
\node[box, fill=lorentzblue!10, right=of wemb] (proj) {HyTE-H\\Projection};
\node[box, fill=lorentzblue!15, right=of proj] (idx) {Lorentz\\Index};

\node[box, fill=bm25orange!10, below=2.0cm of wemb] (bm25) {BM25\\Lexical Index};
\node[box, fill=bm25orange!10, left=of bm25] (query) {User\\Query};
\node[box, fill=bm25orange!10, right=of bm25] (cand) {Candidate\\Set};
\node[box, fill=hybridgreen!10, right=of cand] (rerank) {Lorentz\\Rerank};
\node[box, fill=hybridgreen!10, right=of rerank] (out) {Top-$k$\\Results};

\draw[arrow, lorentzblue] (docs) -- (chunk);
\draw[arrow, lorentzblue] (chunk) -- (wemb);
\draw[arrow, lorentzblue] (wemb) -- (proj);
\draw[arrow, lorentzblue] (proj) -- (idx);

\draw[arrow, bm25orange] (query) -- (bm25);
\draw[arrow, bm25orange] (bm25) -- (cand);
\draw[arrow, hybridgreen] (cand) -- (rerank);
\draw[arrow, hybridgreen] (rerank) -- (out);

\draw[arrow, dashed, lorentzblue] (query.north) to[out=90, in=180] (bm25.north);
\draw[arrow, dashed, hybridgreen] (idx.south) to[out=-60, in=120] (rerank.north);

\node[phase, fit=(docs)(chunk)(wemb)(proj)(idx), label={\textbf{Indexing Phase}}] {};
\node[phase, fit=(query)(bm25)(cand)(rerank)(out), label={\textbf{Retrieval Phase}}] {};

\end{tikzpicture}
}
\caption{System architecture. The indexing phase (top) processes documents through chunking, word embedding lookup, HyTE-H hyperbolic projection, and Lorentz index construction. The retrieval phase (bottom) is two-stage: BM25 lexical search finds candidates, then Lorentz inner-product similarity re-ranks them.}
\label{fig:architecture}
\end{figure*}

Figure~\ref{fig:architecture} shows the full pipeline. The system operates in two distinct phases: an offline \textbf{indexing phase} that processes documents into a persistent Lorentz index, and an online \textbf{retrieval phase} that responds to user queries through a hybrid sparse-dense two-stage pipeline. Both phases run entirely on edge devices.

\subsection{The Lorentz Model of Hyperbolic Geometry}

This system functions in the $d$-dimensional Lorentz hyperboloid model, also known as the Minkowski model, which parameterises the hyperbolic space of constant negative curvature $-1/K$ (where $K > 0$ is the curvature parameter) as the upper sheet of a two-sheeted hyperboloid embedded in $(d+1)$-dimensional Minkowski space $\mathbb{R}^{d+1,1}$:

\begin{equation}
\begin{aligned}
\mathbb{H}^d_K = \Big\{ \mathbf{p} = (t, \mathbf{x}) &\in \mathbb{R}^{1, d} \;:\; \\
&\langle\mathbf{p}, \mathbf{p}\rangle_{\mathcal{L}} = -K,\; t > 0 \Big\}
\end{aligned}
\end{equation}

where the Minkowski inner product is defined as:

\begin{equation}
\begin{aligned}
\langle\mathbf{p}, \mathbf{q}\rangle_{\mathcal{L}} &= -t_p t_q + \mathbf{x}_p^\top \mathbf{x}_q \\
&= -t_p t_q + \sum_{i=1}^{d} x_p^{(i)} x_q^{(i)}
\end{aligned}
\end{equation}

The time coordinate $t$ is not a free parameter but is determined by the space-like coordinates through the hyperboloid constraint:

\begin{equation}
t = \sqrt{\|\mathbf{x}\|^2 + K}
\end{equation}

Any point on the hyperboloid satisfies $\langle\mathbf{p},\mathbf{p}\rangle_{\mathcal{L}} = -K$.

\begin{figure*}[!tb]
\centering
\begin{tikzpicture}[>=Stealth, scale=1.3]

    \draw[thick, fill=blue!3] (0,0) circle (2.5cm);
    
    \foreach \r/\col in {0.6/blue!40, 1.4/blue!60, 2.2/blue!80} {
        \draw[dashed, \col, opacity=0.5] (0,0) circle (\r);
    }
    \node[blue!40, font=\footnotesize] at (0, -0.85) {general};
    \node[blue!60, font=\footnotesize] at (0, -1.65) {domain};
    \node[blue!80, font=\footnotesize] at (0, -2.26) {specific};
    
    \fill[red] (0,0) circle (2pt) node[below right, red, font=\footnotesize] {$\mathbf{0}$};
    
    \fill[lorentzblue] ( 0.50,  0.33) circle (1.5pt) node[font=\tiny, right] {ML};
    \fill[lorentzblue] (-0.40,  0.45) circle (1.5pt) node[font=\tiny, left]  {NLP};
    \fill[lorentzblue] (-0.10, -0.50) circle (1.5pt) node[font=\tiny, below] {CV};
    
    \fill[lorentzblue] ( 1.00, -0.60) circle (1.5pt) node[font=\tiny, right] {GANs};
    \fill[lorentzblue] (-1.10, -0.40) circle (1.5pt) node[font=\tiny, left]  {Transformers};
    \fill[lorentzblue] (-0.60,  1.10) circle (1.5pt) node[font=\tiny, above] {Retrieval};
    
    \fill[lorentzblue] ( 1.65,  0.10) circle (1.5pt) node[font=\tiny, right] {DCGAN};
    \fill[lorentzblue] (-1.84, -0.20) circle (1.5pt) node[font=\tiny, left]  {BERT};
    \fill[lorentzblue] (-1.00, -1.60) circle (1.5pt) node[font=\tiny, below] {HyTE-H};
    
    \draw[red, dashed, thick] (0.50, 0.33) to[out=40, in=160] (1.00, -0.60);
    
    \draw[-{Stealth[length=2mm]}, gray, thick] (0,0) -- (2.0, 1.2) 
        node[midway, above, font=\tiny, rotate=30, gray] {radial depth $r$};

\end{tikzpicture}
\caption{Hyperbolic manifold geometry in the Poincar\'{e} disk model. Concepts organise into hierarchical levels (general, domain, and specific), with radial depth $r = \operatorname{arcosh}(t/\sqrt{K})$ measuring specificity. In the machine learning example shown, broad terms such as \emph{ML} and \emph{NLP} sit near the origin, while more specific terms such as \emph{DCGAN} and \emph{HyTE-H} sit near the boundary. The dashed red curve is a hyperbolic geodesic, the shortest path between two points on the manifold, which curves toward the origin. A Euclidean straight line between the same points would leave the disk.}
\label{fig:manifold}
\end{figure*}
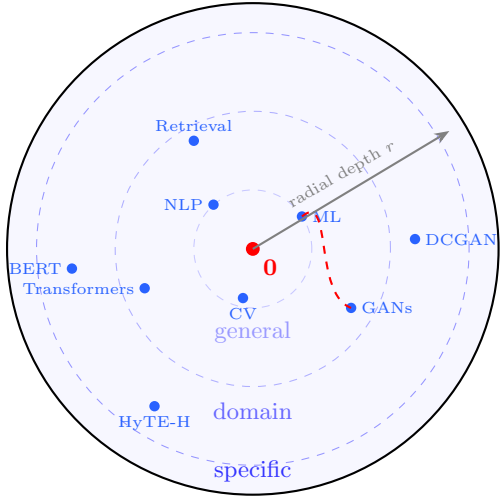

\begin{figure*}[!tb]
\centering
\begin{tikzpicture}[>=Stealth, scale=1.3]

    \draw[->, thick] (-3.2, 0) -- (3.2, 0) node[above left] {$x$ (space)};
    \draw[->, thick] (0, -0.3) -- (0, 3.5) node[above right] {$t$ (time/hierarchy)};

    \fill[red] (0, 1) circle (2.2pt) node[below right, red] {$\mathbf{0}$};

    \draw[blue, thick, domain=-2.4:2.4, samples=100] 
        plot (\x, {sqrt(\x*\x + 1)});

    \draw[blue, dashed, opacity=0.5] (0, 2.6) ellipse (2.4cm and 0.25cm);

    \coordinate (P1) at (-1.0, 1.414);
    \coordinate (P3) at (2.1, 2.326);
    \coordinate (P2) at (0.4, 1.2);

    \fill[blue] (P1) circle (2pt) node[left=3pt, blue] {$p_1$ (general)};
    \fill[blue] (P3) circle (2pt) node[right=3pt, blue] {$p_3$ (specific)};
    \fill[blue] (P2) circle (2pt) node[right=3pt, blue] {$p_2$};

    \draw[red, dashed, thick] (P1) to[out=-20, in=200] (P3);

\end{tikzpicture}

\caption{The Lorentz hyperboloid $\mathbb{H}^1_K$ (one spatial dimension). The time coordinate $t$ encodes hierarchical depth. Points farther from the origin (larger $t$) are more specific concepts. The dashed red curve is a geodesic between two points, as in Figure~\ref{fig:manifold}.}
\label{fig:lorentz}
\end{figure*}
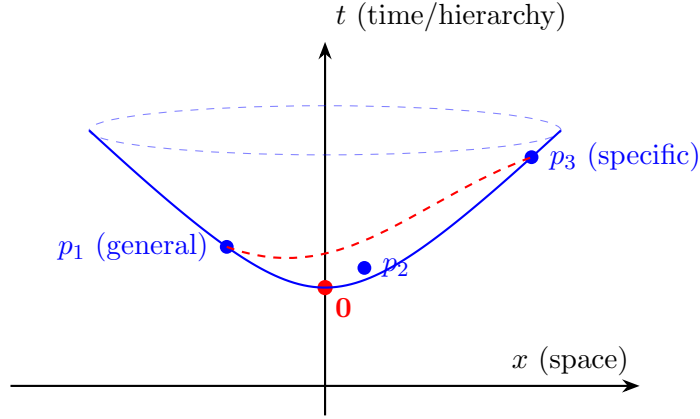

The geodesic distance between two points $\mathbf{p}, \mathbf{q} \in \mathbb{H}^d_K$ is given by:

\begin{equation}
d_{\mathcal{L}}(\mathbf{p}, \mathbf{q}) = \sqrt{K} \cdot \operatorname{arcosh}\left(-\frac{\langle\mathbf{p}, \mathbf{q}\rangle_{\mathcal{L}}}{K}\right)
\end{equation}

Since $\operatorname{arcosh}$ is monotonic, maximising the inner product $\langle\mathbf{p}, \mathbf{q}\rangle_{\mathcal{L}}$ (making it less negative) is equivalent to minimising geodesic distance. The inner product is used directly as the similarity score; the inverse hyperbolic cosine never needs to be computed.

\subsection{Word Embedding and HyTE-H Projection}

Our token-level encoder begins with word embeddings from BGE-small-en-v1.5~\cite{xiao2023cpack}, a 384-dimensional BERT-based model pretrained on massive text pairs for retrieval. The word embedding matrix $\mathbf{W} \in \mathbb{R}^{V \times 384}$ (with vocabulary size $V = 30,522$) is extracted and each row is normalized to unit Euclidean norm. These embeddings are never updated during training or inference; the projection layer only learns to map them into hyperbolic space.

The HyTE-H projection module is a two-layer nonlinear transformation:

\begin{equation}
\begin{aligned}
\mathbf{h} &= \operatorname{ReLU}\left(\mathbf{W}_1 \mathbf{e}\right), \\
\mathbf{p}_{\text{space}} &= \mathbf{W}_2 \mathbf{h}
\end{aligned}
\end{equation}

where $\mathbf{e} \in \mathbb{R}^{384}$ is the input word embedding, $\mathbf{W}_1 \in \mathbb{R}^{256 \times 384}$ projects into a 256-dimensional hidden space, and $\mathbf{W}_2 \in \mathbb{R}^{384 \times 256}$ expands back to the original embedding dimensionality. The output $\mathbf{p}_{\text{space}}$ forms the spatial coordinates of the hyperbolic point; the time coordinate is then computed through the \texttt{add\_time} operation:

\begin{equation}
\begin{aligned}
t &= \sqrt{\|\mathbf{p}_{\text{space}}\|^2 + K}, \\
\mathbf{p} &= (t, \mathbf{p}_{\text{space}}) \in \mathbb{H}^{384}_K
\end{aligned}
\end{equation}

with curvature $K = 1.0$.

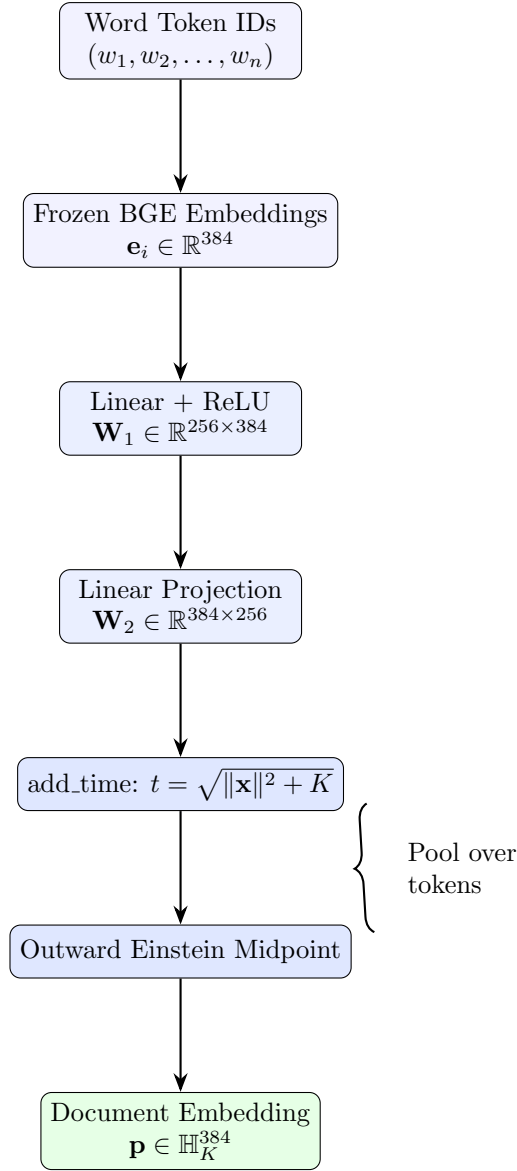
\begin{figure*}[!tb]
\centering
\begin{tikzpicture}[
    node distance=1.5cm,
    layer/.style={rectangle, draw, rounded corners, minimum width=3.2cm, minimum height=0.7cm, align=center, font=\small},
    arrow/.style={-{Stealth[length=2.5mm]}, thick}
]
\node[layer, fill=blue!5] (vocab) {Word Token IDs\\$(w_1, w_2, \ldots, w_n)$};
\node[layer, fill=blue!5, below=of vocab] (lookup) {Frozen BGE Embeddings\\$\mathbf{e}_i \in \mathbb{R}^{384}$};
\node[layer, fill=lorentzblue!10, below=of lookup] (w1) {Linear + ReLU\\$\mathbf{W}_1 \in \mathbb{R}^{256 \times 384}$};
\node[layer, fill=lorentzblue!10, below=of w1] (w2) {Linear Projection\\$\mathbf{W}_2 \in \mathbb{R}^{384 \times 256}$};
\node[layer, fill=lorentzblue!15, below=of w2] (time) {add\_time: $t = \sqrt{\|\mathbf{x}\|^2 + K}$};
\node[layer, fill=lorentzblue!15, below=of time] (oem) {Outward Einstein Midpoint};
\node[layer, fill=green!10, below=of oem] (final) {Document Embedding\\$\mathbf{p} \in \mathbb{H}^{384}_K$};

\draw[arrow] (vocab) -- (lookup);
\draw[arrow] (lookup) -- (w1);
\draw[arrow] (w1) -- (w2);
\draw[arrow] (w2) -- (time);
\draw[arrow] (time) -- (oem);
\draw[arrow] (oem) -- (final);

\draw[decorate, decoration={brace, amplitude=6pt, mirror}, thick] 
    ($(time.south east) + (0.3, 0.1)$) -- ($(oem.north east) + (0.3, -0.1)$)
    node[midway, right=10pt, font=\small, align=left] {Pool over\\tokens};

\end{tikzpicture}
\caption{The HyTE-H encoding pipeline.}
\label{fig:hyteh}
\end{figure*}

\subsubsection{Training the Projection}

The projection weights $\mathbf{W}_1$ and $\mathbf{W}_2$ are trained contrastively on the MS MARCO v1.1 passage retrieval dataset~\cite{bajaj2016msmarco}, about 8.8~million query-passage pairs collected from Bing search logs. Training streams the data in batches of 192 query-document pairs. Queries and passages are tokenized with the HuggingFace \texttt{tokenizers} library and encoded through the HyTE-H pipeline described above.

The training objective is a standard symmetric cross-entropy loss over the pairwise Lorentz similarity matrix. For a batch of $B$ query embeddings $\{\mathbf{q}_i\}_{i=1}^B$ and document embeddings $\{\mathbf{d}_j\}_{j=1}^B$:

\begin{equation}
\begin{aligned}
\mathcal{L} = \frac{1}{2B} \sum_{i=1}^{B} \Bigg[ &-\log \frac{\exp(s_{ii}/\tau)}{\sum_{j=1}^{B} \exp(s_{ij}/\tau)} \\
&-\log \frac{\exp(s_{ii}/\tau)}{\sum_{j=1}^{B} \exp(s_{ji}/\tau)} \Bigg]
\end{aligned}
\end{equation}

where $s_{ij} = \langle\mathbf{q}_i, \mathbf{d}_j\rangle_{\mathcal{L}} / \tau$ is the temperature-scaled Lorentzian inner product with $\tau = 0.05$.

Training runs for 75,000 steps with the AdamW optimiser~\cite{loshchilov2017adamw}, a learning rate of $2 \times 10^{-4}$, and weight decay of 0.01. Training uses a linear warmup over the first 1,000 steps followed by a cosine decay schedule, with automatic mixed precision (AMP) and gradient scaling on two NVIDIA Tesla T4 GPUs (16~GB VRAM). The total number of trainable parameters is 196,608 (approximately 768~KB when stored as 32-bit floats), making the projection module easily deployable on edge devices.

\begin{figure*}[!tb]
\centering

\begin{tikzpicture}[
    node distance=0.6cm and 0.6cm,
    box/.style={rectangle, draw=lorentzblue!80, fill=lorentzblue!5, thick, rounded corners=4pt,
                minimum width=2.0cm, minimum height=0.7cm, align=center, font=\footnotesize},
    docbox/.style={rectangle, draw=bm25orange!80, fill=bm25orange!5, thick, rounded corners=4pt,
                    minimum width=2.0cm, minimum height=0.7cm, align=center, font=\footnotesize},
    mathbox/.style={rectangle, draw=hybridgreen!80, fill=hybridgreen!5, thick, rounded corners=4pt,
                    minimum width=4.0cm, minimum height=0.7cm, align=center, font=\footnotesize},
    lossbox/.style={rectangle, draw=red!60!black, fill=red!3, thick, rounded corners=4pt,
                    minimum width=4.5cm, minimum height=0.8cm, align=center, font=\footnotesize},
    arr/.style={-{Stealth[length=2.5mm]}, thick, draw=gray!60},
    darr/.style={-{Stealth[length=2.5mm]}, thick, dashed, draw=red!60!black}
]

\node[box] (qtok) {Query Tokens};
\node[box, below=of qtok] (qproj) {HyTE-H};
\node[box, below=of qproj] (qe) {$\mathbf{q}_i \in \mathbb{H}^{384}_K$};

\node[docbox, right=2.5cm of qtok] (dtok) {Doc Tokens};
\node[docbox, below=of dtok] (dproj) {HyTE-H};
\node[docbox, below=of dproj] (de) {$\mathbf{d}_j \in \mathbb{H}^{384}_K$};

\node[mathbox, below=1.2cm of $(qe.south)!0.5!(de.south)$] (sim) 
    {$s_{ij} = \langle\mathbf{q}_i, \mathbf{d}_j\rangle_{\mathcal{L}} \;/\; \tau$};

\node[lossbox, below=1.2cm of sim] (loss) {Cross-Entropy Loss $\mathcal{L}$};

\draw[arr] (qtok) -- (qproj);
\draw[arr] (qproj) -- (qe);
\draw[arr] (dtok) -- (dproj);
\draw[arr] (dproj) -- (de);

\draw[arr] (qe.south) -- ++(0,-0.4) -| (sim.north);
\draw[arr] (de.south) -- ++(0,-0.4) -| (sim.north);

\draw[arr] (sim) -- (loss);

\draw[darr] (loss.west) -- ++(-1.8,0) 
    node[midway, above, font=\footnotesize] {$\nabla_{\mathbf{W}}$}
    |- (qproj.west);

\draw[darr] (loss.east) -- ++(1.8,0) 
    node[midway, above, font=\footnotesize] {$\nabla_{\mathbf{W}}$}
    |- (dproj.east);

\end{tikzpicture}

\caption{Contrastive HyTE-H training. Query and document tokens pass through identical HyTE-H modules to project them into a shared hyperbolic space for similarity scoring. The resulting cross-entropy loss updates only the projection weights; the base embeddings stay frozen.}
\label{fig:training}
\end{figure*}
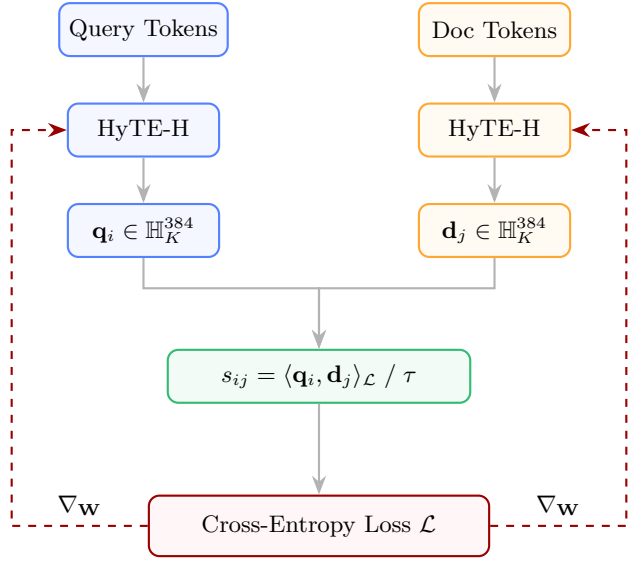

\subsection{Sliding Window Chunking}

Documents are segmented into overlapping chunks using a sliding window with configurable stride and overlap parameters. By default the window is $W = 300$ words, with $O = 60$ words of overlap between consecutive chunks. For a document of length $L$ words, this produces $\lceil L / (W - O) \rceil$ chunks. Each chunk is independently embedded through the HyTE-H pipeline and stored as a separate entry in the Lorentz index, preserving the ability to retrieve at the passage level rather than the document level.

\subsection{BM25 Lexical Index}

\begin{figure*}[!tb]
\centering
\begin{tikzpicture}[
    node distance=1.2cm,
    box/.style={rectangle, draw, fill=bm25orange!8, minimum width=2.5cm, minimum height=0.7cm, align=center, font=\footnotesize, rounded corners},
    arrow/.style={-{Stealth[length=2.5mm]}, thick}
]
\node[box] (doc) {Document $d$};
\node[box, right=of doc] (token) {Tokenize \&\\Normalize};
\node[box, right=of token] (tf) {Term Frequency\\$\text{tf}(t,d)$};
\node[box, below=of tf] (idf) {Inverse Document Frequency\\$\text{idf}(t) = \log\frac{N - n_t + 0.5}{n_t + 0.5}$};
\node[box, left=of idf] (score) {BM25 Score\\$\sum_{t \in q} \text{idf}(t) \cdot \frac{\text{tf}(t,d)(k_1+1)}{\text{tf}(t,d)+k_1}$};

\draw[arrow] (doc) -- (token);
\draw[arrow] (token) -- (tf);
\draw[arrow] (tf) -- (score);
\draw[arrow] (idf) -- (score);
\end{tikzpicture}
\caption{BM25 scoring process.}
\label{fig:bm25}
\end{figure*}
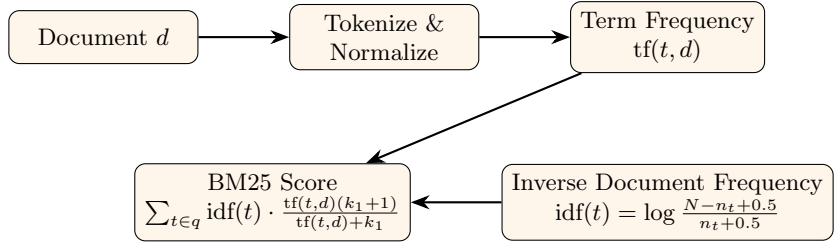

The BM25~\cite{robertson2009probabilistic} implementation uses the standard Okapi formulation with $k_1 = 1.2$ (term frequency saturation) and $b = 0.75$ (length normalisation). Tokens come from whitespace splitting and lowercasing, with no stemming or stop-word removal. For each term, the inverted index stores the documents that contain it and their term frequencies.

\subsection{Two-Stage Retrieval}

Retrieval runs in two stages. First, BM25 scoring over the document corpus produces an initial candidate ranking. Second, for each candidate, its Lorentz embedding is compared against the query embedding (encoded through the same HyTE-H pipeline), and the two scores are linearly interpolated:
\begin{equation}
\text{score}(d, q) = \alpha \cdot s_{\text{bm}}(d,q) + (1 - \alpha) \cdot s_{\mathcal{L}}(d,q)
\end{equation}
where the component scores are
\begin{align}
s_{\text{bm}}(d,q)
  &= \frac{\operatorname{BM25}(d,q)}{\max_{d'}\operatorname{BM25}(d',q)}, \\
s_{\mathcal{L}}(d,q)
  &= \frac{1}{1+\exp\left(-(\langle\mathbf{p}_d,\mathbf{p}_q\rangle_{\mathcal{L}}+2)\right)}.
\end{align}
The first is the normalised lexical score, while the second maps the Lorentz inner product to $[0,1]$ through a logistic function of the raw similarity. The parameter $\alpha \in [0,1]$ sets the weight of each score, and $K=1.0$ is the curvature constant.

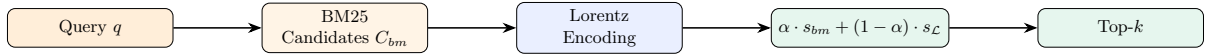
\begin{figure*}[!tb]
\centering
\resizebox{\textwidth}{!}{%
\begin{tikzpicture}[
    node distance=1.6cm,
    box/.style={rectangle, draw, minimum width=3.0cm, minimum height=0.7cm, align=center, font=\footnotesize, rounded corners},
    arr/.style={-{Stealth[length=2.5mm]}, thick}
]
\node[box, fill=bm25orange!15] (q) {Query $q$};
\node[box, fill=bm25orange!10, right=of q] (bm25) {BM25\\Candidates $C_{bm}$};
\node[box, fill=lorentzblue!10, right=of bm25] (enc) {Lorentz\\Encoding};
\node[box, fill=hybridgreen!10, right=of enc] (score) {$\alpha \cdot s_{bm} + (1-\alpha) \cdot s_{\mathcal{L}}$};
\node[box, fill=hybridgreen!10, right=of score] (topk) {Top-$k$};

\draw[arr] (q) -- (bm25);
\draw[arr] (bm25) -- (enc);
\draw[arr] (enc) -- (score);
\draw[arr] (score) -- (topk);
\end{tikzpicture}
}
\caption{Two-stage retrieval with tunable $\alpha$.}
\label{fig:rerank}
\end{figure*}

This two-stage design is motivated by efficiency. Comparing the query against every chunk in the index would cost $O(N \cdot d)$, where $N$ is the number of chunks and $d$ the embedding dimensionality (384). By restricting Lorentz scoring to only the top-$M$ BM25 candidates (with $M = 50$ in practice), the expensive hyperbolic computation is decoupled from corpus size and depends only on the constant $M$. BM25 itself reads the inverted index, so its cost scales with the number of query terms and the postings lists they touch, not with the number of chunks.

\subsection{Persistent Index Serialization}

Since this system is meant to be used on edge devices, the C++ implementation supports persistence of the index. Serialization is implemented through a three-file storage format within the application's private files directory. The \texttt{idx.bin} file stores the Lorentz index as a raw binary dump of the embedding matrix (float32, $N \times (d+1)$, the extra dimension being the time coordinate), the chunk metadata (file path and chunk index for each embedding), and the BM25 inverted index state. The embedding matrix is one contiguous block of floats, one vector per chunk, so scoring walks sequential memory and the file maps directly onto the in-memory array on load. The \texttt{texts.z} file stores the chunk text snippets compressed using zlib at the default compression level, achieving approximately $3$--$5\times$ compression on general texts. The \texttt{paths.txt} file lists the source document paths, one per line, so the system can tell which documents are indexed after a restart.

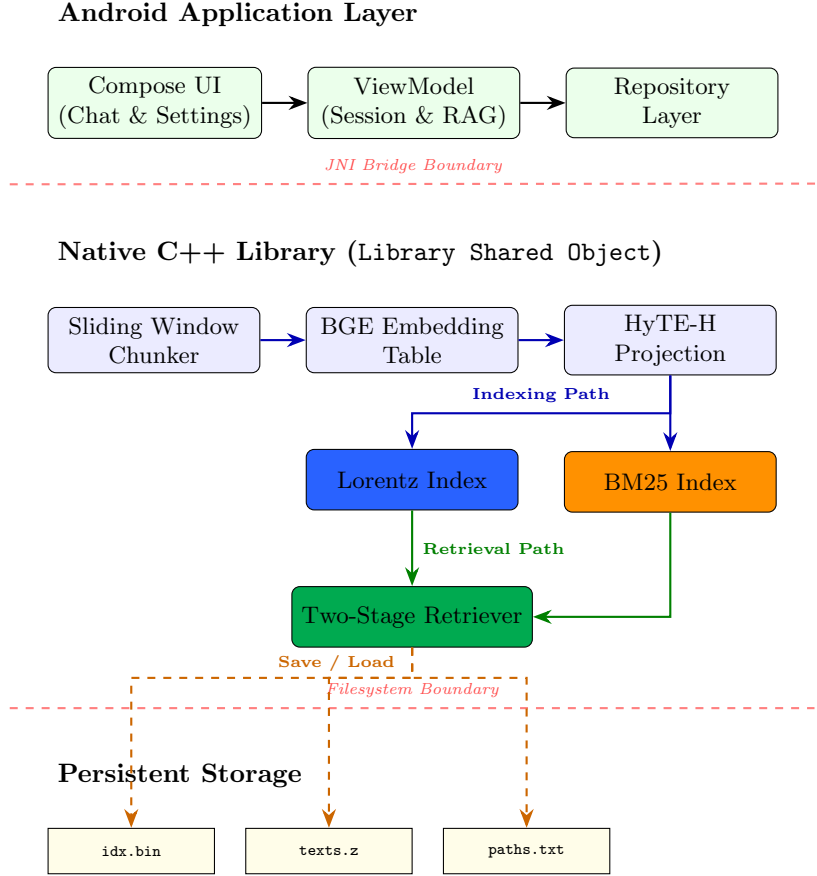
\begin{figure*}[!tb]
\centering

\begin{tikzpicture}[
    node distance=0.8cm and 0.6cm,
    block/.style={rectangle, draw, rounded corners=3pt, minimum width=2.8cm, minimum height=0.8cm, align=center, font=\footnotesize},
    store/.style={rectangle, draw, fill=yellow!10, minimum width=2.2cm, minimum height=0.6cm, align=center, font=\ttfamily\tiny},
    arrow/.style={-{Stealth[length=2.5mm]}, thick},
    bound/.style={draw=red!50, dashed, thick},
    section title/.style={font=\small\bfseries, align=left}
]

\node[section title] (app_title) at (0, 0) {Android Application Layer};

\node[block, fill=green!8, below=0.4cm of app_title.south west, anchor=north west] (ui) {Compose UI\\(Chat \& Settings)};
\node[block, fill=green!8, right=of ui] (vm) {ViewModel\\(Session \& RAG)};
\node[block, fill=green!8, right=of vm] (repo) {Repository\\Layer};

\draw[arrow] (ui) -- (vm);
\draw[arrow] (vm) -- (repo);

\path (ui.south west) ++(-0.5, -0.6) coordinate (jni_left);
\path (repo.south east) ++(0.5, -0.6) coordinate (jni_right);
\draw[bound] (jni_left) -- (jni_right) node[midway, above, font=\tiny\itshape, red!70] {JNI Bridge Boundary};

\path (ui.south west) ++(0, -1.2) coordinate (native_pos);
\node[section title, anchor=north west] (native_title) at (native_pos) {Native C++ Library (\texttt{Library Shared Object})};

\node[block, fill=blue!8, below=0.4cm of native_title.south west, anchor=north west] (chunker) {Sliding Window\\Chunker};
\node[block, fill=blue!8, right=of chunker] (embed) {BGE Embedding\\Table};
\node[block, fill=blue!8, right=of embed] (projnode) {HyTE-H\\Projection};

\node[block, fill=lorentzblue, below=1.0cm of embed] (lidx) {Lorentz Index};
\node[block, fill=bm25orange, below=1.0cm of projnode] (bm25node) {BM25 Index};

\node[block, fill=hybridgreen, below=1.0cm of lidx] (ret) {Two-Stage Retriever};

\draw[arrow, blue!70!black] (chunker) -- (embed);
\draw[arrow, blue!70!black] (embed) -- (projnode);
\draw[arrow, blue!70!black] (projnode.south) -- (bm25node.north);
\draw[arrow, blue!70!black] (projnode.south) -- ++(0,-0.5) -| (lidx.north) node[pos=0.25, above, font=\tiny\bfseries] {Indexing Path};

\draw[arrow, green!50!black] (bm25node.south) |- (ret.east);
\draw[arrow, green!50!black] (lidx.south) -- (ret.north) node[midway, right, font=\tiny\bfseries] {Retrieval Path};

\path (chunker.west |- ret.south) ++(-0.5, -0.8) coordinate (fs_left);
\path (projnode.east |- ret.south) ++(0.5, -0.8) coordinate (fs_right);
\draw[bound] (fs_left) -- (fs_right) node[midway, above, font=\tiny\itshape, red!70] {Filesystem Boundary};

\path (chunker.west |- ret.south) ++(0, -1.4) coordinate (storage_pos);
\node[section title, anchor=north west] (storage_title) at (storage_pos) {Persistent Storage};

\node[store, below=0.4cm of storage_title.south west, anchor=north west] (ib) {idx.bin};
\node[store, right=0.4cm of ib] (tz) {texts.z};
\node[store, right=0.4cm of tz] (pt) {paths.txt};

\path (ret.south) ++(-1.0, -0.2) node[font=\tiny\bfseries, orange!80!black] {Save / Load};
\draw[arrow, orange!80!black, dashed] (ret.south) -- ++(0,-0.4) -| (ib.north);
\draw[arrow, orange!80!black, dashed] (ret.south) -- ++(0,-0.4) -| (tz.north);
\draw[arrow, orange!80!black, dashed] (ret.south) -- ++(0,-0.4) -| (pt.north);

\end{tikzpicture}

\caption{Edge runtime architecture. The Android application layer uses JNI bindings to communicate with the native C++ library, which contains the full retrieval system.}
\label{fig:mobile}
\end{figure*}

\section{Implementation Results}\label{sec:results}

\subsection{Experimental Setup}

The retrieval system was evaluated on five datasets from the BEIR (Benchmarking Information Retrieval) suite~\cite{thakur2021beir}, which provides a standardised evaluation framework for zero-shot information retrieval across diverse domains:

\begin{itemize}[leftmargin=*,itemsep=2pt]
    \item \textbf{SciFact} (5,183 documents, 300 queries): Scientific claim verification from biomedical literature.
    \item \textbf{NFCorpus} (3,633 documents, 323 queries): Biomedical articles from PubMed Central.
    \item \textbf{ArguAna} (8,674 documents, 1,406 queries): Counter-argument retrieval from web debates.
    \item \textbf{SciDocs} (25,669 documents, 1,000 queries): Scientific document recommendation based on citation graphs.
    \item \textbf{FiQA} (57,638 documents, 648 queries): Financial question answering from news and reports.
\end{itemize}

All experiments use the same BGE-small-en-v1.5 word embeddings with HyTE-H projection (curvature $K = 1.0$, OEM power $P = 1.0$) and the default chunking configuration of 500-word windows with 90-word overlap. The benchmark executable (\texttt{beir\_bench.exe}) was compiled with Clang targeting x86-64 using the \texttt{-O3} optimisation flag, and experiments were run on a consumer laptop with 16~GB RAM using 2 OpenMP threads. OpenMP parallelized embedding and chunking across documents during indexing and across queries during evaluation. Queries were encoded independently through the same pipeline and matched against pre-built chunk indexes using exhaustive Lorentz nearest-neighbour search.

Three standard retrieval metrics are evaluated, as implemented by the BEIR framework: \textbf{NDCG@10} (Normalised Discounted Cumulative Gain at rank 10), \textbf{MAP@100} (Mean Average Precision at rank 100), and \textbf{Recall@100}.

\subsection{Results}

\begin{table*}[!tb]
\centering
\setlength{\tabcolsep}{5pt}
\caption{BEIR Benchmark Results: NDCG@10 across five datasets.``$+$sec'' indicates section-titled chunking; ``$+$rerank'' indicates BM25 $\to$ Lorentz two-stage reranking.}
\label{tab:results}
\begin{tabular}{llccccc}
\toprule
\multicolumn{2}{l}{\textbf{Configuration}} & \textbf{SciFact} & \textbf{NFCorpus} & \textbf{ArguAna} & \textbf{SciDocs} & \textbf{FiQA} \\
\midrule
\multirow{5}{*}{NDCG@10}
& $\alpha{=}0.3$                   & 0.6537 & 0.3037 & 0.3422 & 0.1495 & 0.2169 \\
& $\alpha{=}0.7$                   & 0.6537 & 0.3037 & 0.3422 & 0.1495 & 0.2168 \\
& $\alpha{=}0.7\,+\,\text{sec}$        & 0.6537 & 0.3037 & 0.3422 & 0.1495 & 0.2168 \\
& $\alpha{=}0.7\,+\,\text{sec}\,+\,\text{rerank}$ & 0.6536 & 0.3037 & 0.3421 & 0.1494 & 0.2165 \\
& $\alpha{=}1.0$                   & 0.6537 & 0.3037 & 0.3422 & 0.1495 & 0.2168 \\
\midrule
\multicolumn{7}{l}{\footnotesize Note: All NDCG@10 values are essentially tied across configurations within each dataset.}\\
\bottomrule
\end{tabular}
\end{table*}

\begin{table*}[!tb]
\centering
\caption{Retrieval latency and index statistics. (\texttt{ms/q} = milliseconds per query averaged over the full query set).}
\label{tab:latency}
\begin{tabular}{lrrrrc}
\toprule
\textbf{Dataset} & \textbf{Docs} & \textbf{Chunks} & \textbf{Index Size} & \textbf{Index Time} & \textbf{ms/q} \\
\midrule
SciFact  & 5,183  & 5,229  & 8.0~MB  & 2.9~s & 4.3 \\
NFCorpus & 3,633  & 3,664  & 5.7~MB  & 1.9~s & 2.7 \\
ArguAna  & 8,674  & 8,750  & 13.5~MB & 1.5~s & 18.7 \\
SciDocs  & 25,669 & 26,199 & 40.3~MB & 5.1~s & 16.0 \\
FiQA     & 57,638 & 59,216 & 91.2~MB & 9.4~s & 29.0 \\
\bottomrule
\end{tabular}
\end{table*}

Within each dataset, NDCG@10 changes by less than 0.001 across all configurations, regardless of $\alpha$, section-titled chunking, or two-stage reranking. Retrieval quality is therefore limited by the word embeddings, not the ranking mechanism. Since all configurations share the same BGE-small embeddings and HyTE-H projection weights, the Lorentz similarity scores produce near-identical top-$k$ rankings. Setting $\alpha = 1.0$ gives pure BM25 and $\alpha = 0.0$ pure Lorentz retrieval, so $\alpha$ is functional. However, on these datasets the two scoring signals happen to favour the same top-ranked documents. On corpora with heavier keyword dependence, $\alpha$ would show a clearer precision-recall trade-off.

Retrieval latency grows linearly with corpus size, as exhaustive nearest-neighbour search predicts. FiQA, the largest dataset at 59,216 chunks, requires 29~ms per query; index construction time is dominated by embedding computation and is a one-time offline cost. The strong performance on SciFact (NDCG@10 = 0.654) confirms that the HyTE-H projection preserves domain-specific scientific terminology despite using general-domain BGE-small embeddings. The lower scores on SciDocs (0.150) and FiQA (0.217) reflect the difficulty of citation-based recommendation and financial jargon tasks without domain-adapted encoders.

\subsection{Radial Hierarchy Verification}

Following HypRAG, it was tested whether the trained HyTE-H projection produces the expected radial hierarchy. The Lorentz embeddings for four concept pairs at different levels of generality were computed, and the radial coordinate $r = \operatorname{arcosh}(t / \sqrt{K})$ of each term was measured (Table \ref{tab:radial}). Three of the four pairs show the expected behaviour: the specific term sits farther from the origin of the hyperboloid than its general counterpart.

\begin{table*}[!tb]
\centering
\caption{Radial hierarchy verification of HyTE-H embeddings. Positive percentage change indicates that the specific term is encoded further from the origin than its general counterpart.}
\label{tab:radial}
\begin{tabular}{l c c c c}
\toprule
\textbf{General} & \textbf{Specific} & \textbf{Radial Shift} & \textbf{Change} & \textbf{Result} \\
\midrule
animal  & mammal  & $7.46 \to 14.70$ & $+97.1\%$ & \textcolor{green!60!black}{SPECIFIC} \\
mammal  & dog     & $14.70 \to 11.48$ & $-21.9\%$ & \textcolor{red!60!black}{COLLAPSED} \\
science & physics & $4.02 \to 7.50$   & $+86.5\%$ & \textcolor{green!60!black}{SPECIFIC} \\
disease & cancer  & $8.79 \to 9.73$   & $+10.8\%$ & \textcolor{green!60!black}{SPECIFIC} \\
\bottomrule
\end{tabular}
\end{table*}

\subsection{Performance}

The Android implementation packages the entire retrieval pipeline into a shared native library compiled for `arm64-v8a` and `armeabi-v7a` architectures. On a mid-range device such as the Samsung SM-S721, a knowledge space of 1,500 chunks loads its 2.3 MB serialized index in under 10 ms, runs BM25 lexical search in under 1 ms per query, and re-ranks the BM25 candidate set by Lorentz similarity in about 2 ms. The full retrieval takes about 3 ms per query. Users can add new files to the index and use them for RAG queries almost immediately, with good accuracy on the device.

\section{Conclusion}\label{sec:conclusion}

This work presents a hybrid hyperbolic retrieval system for edge devices. It combines frozen BGE-small word embeddings, a trained HyTE-H projection into the Lorentz hyperboloid, BM25 lexical scoring, and tunable Lorentzian reranking. The system achieves competitive zero-shot retrieval on five BEIR datasets using a lightweight learned projection ($<$200~K parameters) without fine-tuned encoders or GPU inference. The contribution is a BM25-for-recall, Lorentz-for-precision pipeline that runs on edge devices at interactive latencies ($\sim$3~ms per query).

Future work has four directions. First, replacing the BGE-small word embeddings with a fine-tuned hyperbolic encoder such as HyTE-FH could lift the embedding quality bottleneck identified in our BEIR results. Second, approximate Lorentz nearest-neighbour search via hyperbolic random projections or graph-based navigation would scale retrieval to million-document corpora at sub-10~ms latency. Third, the radial hierarchy, where distance from the origin encodes specificity, could adapt retrieval depth to query generality and improve precision for broad versus narrow queries. Finally, a hyperbolic LLM such as HELM could be paired with this retriever, so that retrieved context and generated text live in the same geometry.

\section*{Acknowledgement}
The author thanks the authors of HypRAG~\cite{madhu2025hyprag}, whose hyperbolic dense retrieval framework is the basis of this work. The author likewise acknowledges the BM25 framework~\cite{robertson2009probabilistic} and the MS MARCO passage retrieval dataset~\cite{bajaj2016msmarco}, which this work builds on. The complete source code is available in the author's GitHub repository~\cite{zetla2026code} at \url{https://github.com/TimeATronics/Zetla/tree/main/src/zetla/rag}.

\bibliographystyle{IEEEtran}
\bibliography{references}

\end{document}